\documentclass[a4paper,11pt]{article}
\usepackage{jheppub}
\usepackage{xcolor}
\usepackage{lmodern}
\usepackage{amsfonts,amsthm,mathtools}
\usepackage{url}
\usepackage{physics}
\usepackage{comment}
\numberwithin{equation}{section}
\AddToHook{cmd/appendix/after}{\renewcommand{\thesection}{\Alph{section}}}
\newcommand{\bea}{\begin{eqnarray}}
\newcommand{\eea}{\end{eqnarray}}
\def\nn{\nonumber}
\allowdisplaybreaks[1]
\def\be{\begin{equation}}
\def\ee{\end{equation}}

\makeatletter
\gdef\@fpheader{}
\gdef\@preprint{}
\makeatother
\renewcommand{\afterLogoSpace}{\vspace*{-23pt}}
\notoc

\title{Gauge-invariant thermodynamic description of rotating black strings in 5D minimal supergravity via Kaluza-Klein reduction}

\author{Yang Yu,}
\author{Peng-Ju Hu}
\affiliation{Center for Joint Quantum Studies and Department of Physics, School of Science,\\ Tianjin University, Tianjin 300350, China}
\emailAdd{yuy@tju.edu.cn}
\emailAdd{pengjuhu@tju.edu.cn}

\abstract{We develop a gauge-invariant formulation of the thermodynamics of rotating black strings in five-dimensional minimal supergravity through Kaluza–Klein reduction. At the two-derivative level, we consistently reduce the bulk action and the Gibbons–Hawking–York boundary term, carefully accounting for the surface contributions induced by the Chern–Simons coupling. The resulting four-dimensional theory is $\mathcal{N}=2$ supergravity coupled to a single vector multiplet, with prepotential $P=\frac{1}{3\sqrt{3}}\frac{X_1^3}{X_0}$. Employing the Hamiltonian formalism, we establish the quantum statistical relation between the Euclidean on-shell action and the Gibbs free energy. We further compute the leading four-derivative corrections to the thermodynamics of a rotating magnetic black string using the Reall–Santos method. The resulting corrections are in precise agreement with previous results obtained directly for the Kerr–Newman black hole in four-dimensional $\mathcal{N}=2$ supergravity.}

\begin{document}
\maketitle

\clearpage
\tableofcontents
\clearpage

\section{Introduction}
Dimensional reduction provides a systematic framework for relating gravitational theories in different dimensions and comparing their black hole solutions. In Kaluza--Klein theory, components of the higher-dimensional metric involving the string direction give rise to scalar and vector fields in the lower-dimensional theory~\cite{Kaluza:1921tu,Klein:1926tv}, while momentum along this direction manifests itself as an electric charge~\cite{Rasheed:1995zv}. Relations between four- and five-dimensional descriptions have also proved particularly useful in the study of supersymmetric black holes and black rings~\cite{Gaiotto:2005gf,Gaiotto:2005xt,Bena:2005ni}. These connections naturally motivate a systematic comparison of black hole thermodynamics across dimensions.  In particular, rotating charged black strings provide an interesting setting for understanding how higher-dimensional conserved charges and their conjugate potentials are mapped to the thermodynamic variables of the reduced theory.

The Euclidean on-shell action plays a central role in black hole thermodynamics, determining the thermodynamic potential in a specified ensemble~\cite{Gibbons:1976ue}. In the Hamiltonian formulation, the on-shell action can be expressed in terms of conserved charges and their conjugate potentials evaluated at spacetime boundaries~\cite{Regge:1974zd,Brown:1992br}. Since dimensional reduction generally produces total derivative terms, a consistent comparison of thermodynamic quantities across dimensions requires careful treatment of boundary contributions, even though these terms do not affect the local equations of motion. In particular, the Einstein--Hilbert action and the Gibbons--Hawking--York boundary term must be reduced together to preserve a well-defined variational principle~\cite{York:1972sj,Gibbons:1976ue,Saskowski:2024tat}. 

Additional subtleties arise in the presence of Chern--Simons interactions, which are not manifestly gauge invariant and can generate nontrivial surface contributions under gauge transformations. These interactions also modify the definitions of conserved charges~\cite{Marolf:2000cb,Hanaki:2007mb}. For magnetically charged black strings, gauge-compensating transgression terms have been introduced to cancel the gauge-dependent contributions induced by Chern--Simons couplings~\cite{Mora:2006ka,Bobev:2026jqd}. Related issues arise from mixed gauge--gravitational Chern--Simons terms in the computation of black string entropy~\cite{Tachikawa:2006sz,Ma:2022nwq}. These observations raise an important question: how can one formulate a gauge-invariant description of black string thermodynamics that satisfies the quantum statistical relation, $I_{\text{on-shell}}=\beta G_{\rm free}$, where $\beta$ denotes the inverse temperature and $G_{\rm free}$ the free energy in the specified ensemble? In this work, we address this question through dimensional reduction. We show that, upon dimensional reduction and extraction of a gauge-noninvariant total derivative term, the 5D action reduces to a fully gauge-invariant 4D action. Based on Hamiltonian formalism of the 4D theory, we establish a gauge-invariant description of the thermodynamic quantities and the quantum statistical relation. Thus our approach provides an alternative to the aforementioned transgression completion prescription.

Higher-derivative interactions provide a further motivation for our analysis, as they naturally arise in effective theories of quantum gravity and modify black hole thermodynamics~\cite{Wald:1993nt,Iyer:1994ys}. We consider the curvature-squared extensions of five-dimensional minimal supergravity constructed using superconformal tensor calculus~\cite{Hanaki:2006pj,Ozkan:2013uk,Ozkan:2013nwa}. For asymptotically flat solutions, thermodynamic quantities are invariant under perturbative field redefinitions~\cite{Ma:2023qqj}, allowing the higher-derivative action to be simplified before evaluating the on-shell action. In particular, the three curvature-squared supergravity invariants can be reduced, up to field redefinitions, to a single independent combination containing both the Weyl-tensor-squared and mixed gauge--gravitational Chern--Simons couplings~\cite{Gold:2023ymc,Ozkan:2024euj,Liu:2022sew,Ma:2024ynp,Cai:2026qrr}. Similar simplifications have also been explored in four-dimensional $\mathcal{N}=2$ supergravity~\cite{Bobev:2021oku,Hu:2023qhs,Hu:2025ogz}. To evaluate the leading higher-derivative corrections, we employ the Reall--Santos method~\cite{Reall:2019sah}, which determines the leading-order correction to the Euclidean on-shell action by evaluating the higher-derivative interactions on the uncorrected background, subject to appropriate boundary conditions. This method and its extensions~\cite{Ma:2023qqj,Hu:2023gru,Xiao:2023two,Zhang:2026knn} are particularly advantageous for rotating solutions, for which constructing the corrected geometry can be technically challenging~\cite{Hu:2025ogz,Hu:2025aji,Ma:2024ynp,Chen:2024knw}. In this context, the surface terms generated by dimensional reduction deserve special attention, as they can contribute to the higher-derivative corrections to the on-shell action.

As a concrete application, we consider the rotating charged black string solution of five-dimensional minimal supergravity~\cite{Compere:2010fm}. Unlike five-dimensional black holes, black strings admit a dimensional reduction to asymptotically flat four-dimensional black holes, making them particularly suitable for our analysis. For simplicity, we focus on the special case corresponding to uplifting the four-dimensional Kerr--Newman black hole. This allows us to exploit the well-understood thermodynamics of Kerr--Newman black holes while investigating the higher-derivative corrections and the role of surface terms in relating the four- and five-dimensional on-shell actions. We show that our results are in agreement with those previously obtained directly from the four-dimensional description.

The remainder of this paper is organized as follows. In Sec.~\ref{sec:supergravity-setup}, we review the general rotating black string solution and its two-derivative thermodynamics. In Sec.~\ref{sec:KK-reduction}, we perform the dimensional reduction of the bulk and boundary actions. In Sec.~\ref{sec:thermodynamic-hamiltonian}, we develop the Hamiltonian formulation and establish the quantum statistical relation. In Sec.~\ref{sec:higher-derivative-corrections}, we evaluate the leading four-derivative corrections to the on-shell action and compare the four- and five-dimensional results. We conclude in Sec.~\ref{sec:discussion}.

Throughout the paper, we adopt the metric signature $(-,+,+,+,+)$. Greek letters $\alpha,\beta,\gamma,\ldots$ and uppercase Latin letters $A,B,C,\ldots$ denote five-dimensional spacetime and Lorentz indices, respectively, while Greek letters $\mu,\nu,\rho,\ldots$ and lowercase Latin letters $a,b,c,\ldots$ denote their four-dimensional counterparts. The indices $i,j,k,\ldots$ label three-dimensional spatial directions. Hatted quantities refer to five-dimensional fields and operators.


\section{Brief review of general black string in 5D minimal supergravity}\label{sec:supergravity-setup}
In this section, we introduce 5D Einstein--Maxwell--Chern--Simons (EMCS) theory, present the rotating black string solution of Ref.~\cite{Compere:2010fm}, and review its thermodynamics.

\subsection{Einstein--Maxwell--Chern--Simons theory and the black string solution}
The bosonic sector of 5D minimal supergravity is
described by EMCS theory. Including the gravitational boundary term at spatial
infinity, the action reads
\begin{align}
I^{(5)} &= I_{\rm EMCS}^{(5)}+I_{\rm GHY}^{(5)},
\label{eq:5d-two-derivative-action}
\end{align}
where
\begin{align}
I_{\rm EMCS}^{(5)}
&= \frac{1}{16\pi G_{5}}
\int_{\mathcal{M}_5}\mathrm{d}^{5}x\,\sqrt{-\hat{g}}
\left(
\hat{R}-\frac{1}{4}\hat{F}^{2}
-\frac{1}{12\sqrt{3}}
\hat{\epsilon}^{\alpha\beta\gamma\delta\zeta}
\hat{F}_{\alpha\beta}\hat{F}_{\gamma\delta}\hat{A}_{\zeta}
\right), \label{eq:5DEMCS}
\\
I_{\rm GHY}^{(5)}
&= \frac{1}{8\pi G_{5}}
\int_{\partial\mathcal{M}_{5}}
\mathrm{d}^{4}x\,\sqrt{-\hat l}
\left(K[\hat l]-K[\hat{\bar{l}}]\right).\label{eq:5D-GHY}
\end{align}
Here $G_{5}$ is the 5D Newton constant,
$\hat{F}=\mathrm{d}\hat{A}$, $\hat{F}^{2}=\hat{F}_{\alpha\beta}\hat{F}^{\alpha\beta}$, and
$\hat{\epsilon}^{\alpha\beta\gamma\delta\zeta}$ is the 5D Levi--Civita tensor. The asymptotic boundary $\partial\mathcal{M}_{5}$ is understood as the limit of a timelike hypersurface at spatial infinity. Its outward-pointing unit normal $\hat{r}^{\alpha}$ is spacelike, with $\hat r_{\alpha}\hat r^{\alpha}=1$, and the induced metric is given by
\begin{equation}
    \hat l_{\alpha\beta}=\hat{g}_{\alpha\beta}-\hat r_{\alpha}\hat r_{\beta},
    \label{eq:imetric-l}
\end{equation}
whose pullback to the boundary is denoted by $l_{\mu\nu}$. The extrinsic curvature is defined as
\begin{align}
K_{\alpha\beta}[\hat{l}]&\coloneqq \hat{l}_{\alpha}{}^{\gamma}\hat{\nabla}_{\gamma}\hat r_{\beta}.
\end{align}
The Gibbons--Hawking--York (GHY) term ensures a well-posed Dirichlet variational principle for the gravitational sector~\cite{York:1972sj,Gibbons:1976ue}. The reference term $K[\bar l]$ is the trace of the extrinsic curvature of the corresponding boundary embedded in a reference background, with the induced metric $\bar l_{\alpha\beta}$ matched to $l_{\alpha\beta}$ at the cutoff surface. This term implements background subtraction.

We consider the stationary, rotating black string family constructed in Ref.~\cite{Compere:2010fm}, which depends on five parameters $(m,a,\beta,\gamma,\delta)$. Here $m$ denotes the mass parameter, $a$ controls rotation in the transverse space, $\gamma$ parametrizes a boost along the string direction, and $\beta$ and $\delta$ control the magnetic and electric charges, respectively. We use coordinates $(t,r,x,\phi,z)$, where $z$ is the direction along which the string extends. We introduce the shorthand notation
\begin{align}
s_{X}&\coloneqq \sinh X,\quad\quad c_{X}\coloneqq \cosh X.
\end{align}
The metric and gauge potential take the form
\begin{align}
&\mathrm{d}s_5^{2}  =(F_{1}+\Delta_{2})\big(\mathrm{d}s_{\text{base}}^{2}-\frac{\Delta_{2}}{\xi}(\mathrm{d}t+\omega_{\phi}\mathrm{d}\phi)^{2}\big)+\frac{\xi}{(F_{1}+\Delta_{2})^{2}}(\mathrm{d}z+\check{A}_{t}\mathrm{d}t+\check{A}_{\phi}\mathrm{d}\phi)^{2},\nonumber \\
&\hat{A} =\hat{A}_{t}\mathrm{d}t+\hat{A}_{z}\mathrm{d}z+\hat{A}_{\phi}\mathrm{d}\phi,\label{eq:black-string-solution}
\end{align}
with
\begin{align}
\mathrm{d}s_{\text{base}}^{2} & =\big(\frac{\mathrm{d}r^{2}}{\Delta}+\frac{\mathrm{d}x^{2}}{1-x^{2}}\big)+\frac{\Delta(1-x^{2})}{\Delta_{2}}\mathrm{d}\phi^{2},\ \Delta=r^{2}-2mr+a^{2},\ \Delta_{2}=r^{2}-2mr+a^{2}x^{2},\nonumber \\
\hat{A}_{\phi} & =\frac{2\sqrt{3}mc_{\delta}^{2}}{\Delta_{2}}\Big(-\frac{2s_{\beta}c_{\beta}\Delta x}{\sqrt{1+3c_{\delta}^{2}}}-a(1-x^{2})s_{\delta}\big(2ms_{\beta}^{2}(\frac{6c_{\beta}^{2}c_{\delta}^{2}}{1+3c_{\delta}^{2}}-1)+r(c_{\beta}^{2}+s_{\beta}^{2})\big)\Big)\nonumber \\
 & -\sqrt{3}\bar{A}_{\phi}\frac{F_{2}}{(F_{1}+\Delta_{2})}-\sqrt{3}\omega_{\phi}\frac{F_{3}}{(F_{1}+\Delta_{2})},\nonumber \\
\hat{A}_{t} & =-\sqrt{3}\frac{F_{3}}{(F_{1}+\Delta_{2})},\ \ \hat{A}_{z}=-\sqrt{3}\frac{F_{2}}{(F_{1}+\Delta_{2})}.
\end{align}
The auxiliary functions appearing in the metric are
\begin{align}
\xi & =(F_{4}+\Delta_{2})(F_{1}+\Delta_{2})-F_{2}^{2},\nonumber \\
\omega_{\phi} & =-2am\frac{1-x^{2}}{\Delta_{2}}c_{\delta}^{3}\Big(2ms_{\beta}^{2}(c_{\beta}^{2}+s_{\beta}^{2}-\frac{4c_{\beta}^{2}}{1+3c_{\delta}^{2}})+r(c_{\beta}^{2}+s_{\beta}^{2})\Big),\nonumber \\
\check{A}_{\phi} & =\bar{A}_{\phi}+\frac{k}{\xi}\omega_{\phi},\ \ \check{A}_{t}=\frac{k}{\xi},\ \ k=F_{5}(F_{1}+\Delta_{2})-F_{2}F_{3},\cr
\bar{A}_{\phi}&=\frac{4am(1-x^{2})s_{\beta}c_{\beta}c_{\delta}^{3}(r+2s_{\beta}^{2}m)}{\Delta_{2}\sqrt{1+3c_{\delta}^{2}}}.
\label{eq:black-string-functions}
\end{align}
The five functions $F_1,\ldots,F_5$ depend linearly on $r$ and $x$ and are given by
\begin{align}
F_{1} & =2mc_{\delta}^{2}\Big(2ms_{\beta}^{2}(s_{\beta}^{2}+\frac{s_{\delta}^{2}c_{\beta}^{2}}{1+3c_{\delta}^{2}})+r(c_{\beta}^{2}+s_{\beta}^{2})-\frac{2axs_{\delta}s_{\beta}c_{\beta}}{\sqrt{1+3c_{\delta}^{2}}}\Big),\nonumber \\
F_{2} & =-2mc_{\delta}s_{\delta}\Big(\frac{2ms_{\beta}c_{\beta}}{\sqrt{1+3c_{\delta}^{2}}}(-1+(c_{\beta}^{2}+s_{\beta}^{2})c_{\delta}^{2})+\frac{2rs_{\beta}c_{\beta}}{\sqrt{1+3c_{\delta}^{2}}}-axs_{\delta}(c_{\beta}^{2}+s_{\beta}^{2})\Big),\nonumber \\
F_{3} & =2mc_{\delta}\Big(-2ms_{\delta}s_{\beta}^{2}(\frac{c_{\beta}^{2}(1+c_{\delta}^{2})}{1+3c_{\delta}^{2}}-s_{\beta}^{2})+r(c_{\beta}^{2}+s_{\beta}^{2})s_{\delta}+\frac{2axs_{\beta}c_{\beta}}{\sqrt{1+3c_{\delta}^{2}}}\Big),\nonumber \\
F_{4} & =2m\Big(2m\big((c_{\beta}^{2}s_{\delta}^{2}+s_{\beta}^{2}c_{\delta}^{2})^{2}+\frac{s_{\delta}^{2}s_{\beta}^{2}c_{\beta}^{2}}{1+3c_{\delta}^{2}}\big)+r(c_{\beta}^{2}+s_{\beta}^{2})(c_{\delta}^{2}+s_{\delta}^{2})+\frac{2axs_{\delta}s_{\beta}c_{\beta}}{\sqrt{1+3c_{\delta}^{2}}}\Big),\nonumber \\
F_{5} & =2m\Big(\frac{2ms_{\beta}c_{\beta}}{\sqrt{1+3c_{\delta}^{2}}}\left((c_{\beta}^{2}+s_{\beta}^{2})c_{\delta}^{4}-1\right)+\frac{2rs_{\beta}c_{\beta}}{\sqrt{1+3c_{\delta}^{2}}}+axs_{\delta}^{3}(c_{\beta}^{2}+s_{\beta}^{2})\Big).
\label{eq:F1F2F3F4F5}
\end{align}
The parameter $\gamma$ can be introduced by performing a Lorentz boost along the $z$ direction,
\begin{equation}
t\rightarrow c_\gamma\,t-s_\gamma\,z,
\qquad
z\rightarrow c_\gamma\,z-s_\gamma\,t.
\end{equation}
Compared with the parametrization in Ref.~\cite{Compere:2010fm}, we redefine $\delta\to-\delta$ so that the electric charge is positive for $\delta>0$ in our conventions. All subsequent expressions are written in terms of this redefined parameter.

\subsection{Thermodynamics of the general black string in 5D}
We begin by reviewing the two-derivative thermodynamics of the general 5D black string~\cite{Compere:2010fm}. These results provide a reference for the 4D Hamiltonian analysis. To determine the horizon radii of the black string, we write the 5D metric in ADM form \cite{Arnowitt:1962hi,Gourgoulhon:2007ue},
\begin{equation}
\mathrm{d}\hat{s}^{2}_5
=-\hat{N}^{2}\mathrm{d}t^{2}
+\hat{h}_{\mu\nu}
(\mathrm{d}x^{\mu}+\hat{N}^{\mu}\mathrm{d}t)
(\mathrm{d}x^{\nu}+\hat{N}^{\nu}\mathrm{d}t),
\end{equation}
where $\hat{N}$ and $\hat{N}^{\mu}$ are the 5D lapse function and shift vector, respectively. For the stationary foliation of the solution \eqref{eq:black-string-solution}, the horizons are located at the zeros of $\hat{N}^{2}$ \cite{Medved:2004tp}. In terms of the metric components, this condition reads
\begin{equation}
\frac{1}{\hat{g}_{\phi\phi}}
\Big(
\hat{g}^{2}_{t\phi}-\hat{g}_{tt}\hat{g}_{\phi\phi}
-\frac{
4\hat{g}_{xx}
(\hat{g}_{tz}\hat{g}_{\phi\phi}
-\hat{g}_{t\phi}\hat{g}_{\phi z})^{2}
}{
\hat{g}^{2}_{zx}\hat{g}_{\phi\phi}
+4\hat{g}_{xx}
(\hat{g}^{2}_{\phi z}-\hat{g}_{zz}\hat{g}_{\phi\phi})
}
\Big)\Big|_{r=r_{\rm h}}=0.
\end{equation}
Substituting the solution reduces this condition to
\begin{equation}
\Delta
=r_{\rm h}^{2}-2mr_{\rm h}+a^{2}=0.
\end{equation}
The inner and outer horizon radii are therefore
\begin{equation}
r_{-}=m-\sqrt{m^{2}-a^{2}},
\qquad
r_{+}=m+\sqrt{m^{2}-a^{2}},
\end{equation}
respectively, with real roots requiring $m^{2}\geq a^{2}$.
In the following, $r_{\rm h}$ denotes either $r_{-}$ or $r_{+}$.
For $r_{\rm h}\neq0$, the horizon equation allows us to express
$m$ in terms of $r_{\rm h}$ and $a$ as
\begin{equation}
m=\frac{a^{2}+r_{\rm h}^{2}}{2r_{\rm h}}.
\end{equation}

The presence of magnetic charge of the 5D black string requires the gauge potential to be defined on northern and southern patches. Moreover, gauge transformations of the Chern--Simons term can generate surface contributions, so the gauge choice must be specified when evaluating the action. We shift the azimuthal component by $\hat A_\phi\to\hat A_\phi+\hat c_\phi$. At spatial infinity, regularity at the respective poles fixes the gauge constant as
\begin{align}
    \hat A_{\phi}\big|_{r=\infty}
    &= -\Lambda x+\hat c_{\phi}+\mathcal{O}(r^{-1}),
    \qquad
    \hat c_{\phi,\mathrm N}=\Lambda,
    \qquad
    \hat c_{\phi,\mathrm S}=-\Lambda,
\end{align}
where $\Lambda=\frac{2G_5}{\pi R}Q_{\mathrm M}^{(5)}$, with $Q_{\mathrm M}^{(5)}$ denoting the magnetic charge defined in \eqref{eq:5D-thermodynamics-a}. We also shift the temporal component by $\hat A_t\to\hat A_t+\hat c_t$, choosing
\begin{align}
    \hat c_{t,\mathrm N}
    &=-\Omega_{\phi}^{(5)}\Lambda,
    \qquad
    \hat c_{t,\mathrm S}
    =\Omega_{\phi}^{(5)}\Lambda,
\end{align}
so that the contraction of the gauge field $\hat A_\alpha$ with the Killing vector $\hat\xi^\alpha=\hat\partial_{t}+\Omega^{(5)}_{\phi}\hat\partial_{\phi}+v^{(5)}_{z}\hat\partial_z$ is continuous across the patch interface, where $\Omega_{\phi}^{(5)}$ is the angular velocity and $v_{z}^{(5)}$ is the linear velocity along the string direction.

The conserved charges and entropy, parametrized by $(r_{\rm h},a,\beta,\gamma,\delta)$ have been computed in \cite{Compere:2010fm} and they take the form \footnote{Compared to the solution used in~\cite{Compere:2010fm}, we have redefined the parameter $\delta\to-\delta$.}\footnote{Here the string is infinitely long. However, to regularize the integration over $z$ direction, we denote $\int dz=2\pi R$.}
\begin{align}
M^{(5)}&=\frac{2\pi Rm}{G_{5}}G[c_{\gamma},s_{\gamma}],\ \ S^{(5)}=\frac{2\pi^{2}R}{G_{5}}\,F[c_{\gamma},s_{\gamma}],\ \ J_{\phi}^{(5)}=-\frac{2\pi Rmac_{\delta}^{2}}{G_{5}}H[c_{\gamma},-s_{\gamma}],\nonumber\\
P_{z}^{(5)}&=\frac{m\pi R}{G_{5}}\Big[-\frac{2s_{\beta}c_{\beta}(s_{\gamma}^{2}+c_{\gamma}^{2})}{\sqrt{1+3c_{\delta}^{2}}}+c_{\gamma}s_{\gamma}(c_{\beta}^{2}+s_{\beta}^{2})(1+3s_{\delta}^{2})\Big],\nonumber\\
Q_{\text{E}}^{(5)}&=\frac{\sqrt{3}m\pi Rs_{\delta}}{G_{5}}H[c_{\gamma},s_{\gamma}],\quad\quad Q_{\text{M}}^{(5)}=\frac{2\pi R}{16\pi G_{5}}\int_{S^{2}}\hat{F}=\frac{2\pi R\sqrt{3}ms_{\beta}c_{\beta}c_{\delta}^{2}}{G_{5}\sqrt{1+3c_{\delta}^{2}}},
\label{eq:5D-thermodynamics-a}
\end{align}
and the corresponding thermodynamic potentials take the form
\begin{align}
T^{(5)}&=\frac{r_{{\rm h}}^{2}-a^{2}}{4\pi r_{{\rm h}}F[c_{\gamma},s_{\gamma}]},\ \ \ \Omega_{\phi}^{(5)}=-\frac{a}{F[c_{\gamma},s_{\gamma}]},\ \ \ v_{z}^{(5)}=\frac{F[s_{\gamma},c_{\gamma}]}{F[c_{\gamma},s_{\gamma}]},\nonumber\\
\Phi_{\text{E}}^{(5)}&=\hat{\xi}{}^{\alpha}\hat{A}{}_{\alpha}|_{r_{{\rm h}}}^{\infty}=\frac{2\sqrt{3}mc_{\delta}^{2}s_{\delta}}{F[c_{\gamma},s_{\gamma}]}\Big(r_{{\rm h}}c_{\beta}^{2}(\frac{6c_{\delta}^{2}s_{\beta}^{2}}{1+3c_{\delta}^{2}}+1)+\frac{a^{2}}{r_{{\rm h}}}s_{\beta}^{2}(\frac{6c_{\delta}^{2}c_{\beta}^{2}}{1+3c_{\delta}^{2}}-1)\Big),\nonumber\\
\Phi_{\text{M}}^{(5)}&=\frac{2\sqrt{3}\,mc_{\delta}}{F[c_{\gamma},s_{\gamma}]}\bigg[\frac{c_{\beta}s_{\beta}c_{\gamma}}{\sqrt{1+3c_{\delta}^{2}}}\Big(\frac{r_{{\rm h}}^{2}-a^{2}}{r_{{\rm h}}}+2mc_{\delta}^{2}(c_{\beta}^{2}+s_{\beta}^{2})\Big)\nonumber\\
&\qquad-s_{\gamma}\Big(s_{\delta}^{2}\big(\frac{c_{\beta}^{2}r_{{\rm h}}^{2}-a^{2}s_{\beta}^{2}}{r_{{\rm h}}}\big)+\frac{4mc_{\beta}^{2}s_{\beta}^{2}c_{\delta}^{2}(3c_{\delta}^{2}-1)}{1+3c_{\delta}^{2}}\Big)\bigg],
\label{eq:5D-thermodynamics-b}
\end{align}
where the three functions $G[X,Y],\ H[X,Y],\ F[X,Y]$ are defined by
\begin{align}
G[X,Y] & =\frac{-2s_{\beta}c_{\beta}XY}{\sqrt{1+3c_{\delta}^{2}}}+\frac{1}{4}(c_{\beta}^{2}+s_{\beta}^{2})\big(X^{2}-2Y^{2}+3X^{2}(c_{\delta}^{2}+s_{\delta}^{2})\big), \\
H[X,Y] & =c_{\delta}\Big(\frac{2s_{\beta}c_{\beta}Y}{\sqrt{1+3c_{\delta}^{2}}}+(c_{\beta}^{2}+s_{\beta}^{2})X\Big), \\
F[X,Y] & =2mc_{\delta}^{3}\Big[\Big(\frac{c_{\beta}^{4}r^2_{\rm h}+a^{2}s_{\beta}^{4}}{r_{\rm h}}+\frac{6ms_{\beta}^{2}c_{\beta}^{2}s_{\delta}^{2}}{1+3c_{\delta}^{2}}\Big)X-\frac{2s_{\beta}c_{\beta}(c_{\beta}^{2}r_{\rm h}+s_{\beta}^{2}\frac{a^{2}}{r_{\rm h}})}{\sqrt{1+3c_{\delta}^{2}}}Y\Big].
\end{align}
Here $M^{(5)}$ is the ADM mass, $S^{(5)}$ is the entropy, $T^{(5)}$ is the temperature, $\Omega_{\phi}^{(5)}$ is the angular velocity, $J_{\phi}^{(5)}$ is the angular momentum, $v_{z}^{(5)}$ is the linear velocity along the string direction, $P_{z}^{(5)}$ is the corresponding linear momentum, $Q_{\text{E}}^{(5)}$ is the electric charge, $Q_{\text{M}}^{(5)}$ is the magnetic charge and $\Phi_{\text{M}}^{(5)}$ is the magnetic potential.

The mixed ensemble free energy is
\begin{equation}
    G_{\rm free}^{(5)} (T^{(5)},\Phi_{\rm E}^{(5)},\Omega_{\phi}^{(5)},v_{z}^{(5)},Q_{\rm M}^{(5)})=M^{(5)}-T^{(5)} S^{(5)}-\Phi_{\rm E}^{(5)} Q_{\rm E}^{(5)}-\Omega_{\phi}^{(5)}J_{\phi}^{(5)}-v_{z}^{(5)}P_{z}^{(5)},
\end{equation}
from which the grand--canonical free energy follows through the appropriate Legendre transformation
\begin{equation}
    \tilde{G}_{\rm free}^{(5)}(T^{(5)},\Phi_{\rm E}^{(5)},\Omega_{\phi}^{(5)},v_{z}^{(5)},\Phi_{\rm M}^{(5)})=G_{\rm free}^{(5)}-\Phi_{\rm M}^{(5)} Q_{\rm M}^{(5)}.
\end{equation}

For the five-dimensional theory considered here, however, the identification of the free energy with the Euclidean on-shell action multiplied by the temperature has not been established in previous work \cite{Compere:2010fm}. Due to the presence of the bulk Chern--Simons term, an additional boundary contribution is expected to be necessary to restore gauge invariance and establish the quantum statistical relation. This subtlety is also reflected in the dimensional-reduction formula \eqref{eq:action-reduction}, where a residual gauge-noninvariant surface term appears explicitly.

\section{Dimensional reduction and 5D/4D connection}\label{sec:KK-reduction}
We now perform the dimensional reduction of the two-derivative theory along the string direction, retaining also the surface terms needed to match the 5D and 4D actions. The resulting 4D action is fully gauge invariant and can be identified as the ${\cal N}=2$ supergravity coupled to a single vector multiplet with the prepotential $P=\frac1{3\sqrt 3}\frac{X^3_1}{X_0}$. We then exhibit how to further truncate the theory to 4D minimal supergravity.

\subsection{Reduction of the bulk and boundary actions}
We perform a Kaluza--Klein reduction along the string direction $z$, assuming that all fields are independent of $z$~\cite{Compere:2010fm,Kim:2010bf}. Let $e^{a}$ denote an orthonormal coframe for the 4D metric. The reduction ansatz is
\begin{align}
\mathrm{d}s_{5}^{2}
&= e^{\frac{\varphi}{\sqrt{3}}}\,\eta_{ab}e^{a}e^{b}
+e^{-\frac{2\varphi}{\sqrt{3}}}(\hat{e}^{z})^{2},\quad \hat{e}^{z}= \mathrm{d}z+A_{1},\quad \hat{A}=A_{2}+\chi\,\hat{e}^{z}.\label{eq:kk-ansatz}
\end{align}
Here $\varphi$ and $\chi$ are 4D scalar fields, while $A_{1}$ and $ A_{2}$ are one-form gauge potentials. Their field strengths are $F_1=\mathrm{d}A_1$ and $F_2= \mathrm{d}A_2$. Note that this should not be confused with the $F_1,\, F_2$ in Eq.\eqref{eq:F1F2F3F4F5}. The metric $\mathrm{d}s_{4}^{2}=\eta_{ab}e^{a}e^{b}=g_{\mu\nu}\text{d}x^\mu\text{d}x^\nu$ is in the Einstein frame. We denote the Hodge duals associated with the 5D and 4D metrics by
$\star_{5}$ and $\star_{4}$, respectively, and choose compatible orientations such that
\begin{align}
\star_{5}\boldsymbol{1}
&= e^{\frac{\varphi}{\sqrt{3}}}
\star_{4}\boldsymbol{1}\wedge\mathrm{d}z.
\end{align}
In differential-form notation, the 5D bulk action is
\begin{align}
I_{\rm EMCS}^{(5)}
&= \frac{1}{16\pi G_{5}}
\int_{\mathcal{M}_{5}}
\left(
\hat{R}\star_{5}\boldsymbol{1}
-\frac{1}{2}\star_{5}\hat{F}\wedge \hat{F}
+\frac{1}{3\sqrt{3}}\hat{F}\wedge \hat{F}\wedge \hat{A}
\right),
\label{eq:5d-emcs-action}
\end{align}
where $\hat F=\mathrm{d}\hat A$. Substituting the reduction ansatz we obtain
\begin{align}
I_{\rm EMCS}^{(5)}
&= \frac{1}{16\pi G_{5}}
\int_{\mathcal{M}_5}
\mathcal{L}^{(4)}\wedge\mathrm{d}z
+I_{\rm B},
\label{eq:reduced-emcs-action}
\end{align}
where
\begin{align}
\mathcal{L}^{(4)}
=& \Big[
R
-\frac{1}{2}(\nabla\varphi)^{2}
-\frac{1}{2}e^{\frac{2\varphi}{\sqrt{3}}}(\nabla\chi)^{2}
-\frac{1}{4}e^{-\sqrt{3}\varphi}F_{1}^{2}
-\frac{1}{4}e^{-\frac{\varphi}{\sqrt{3}}}(F_{2}+\chi F_{1})^{2}
\Big]\star_{4}\boldsymbol{1}
\nonumber\\
&+\frac{1}{\sqrt{3}}
\Big(
\chi F_{2}\wedge F_{2}
+\chi^{2}F_{1}\wedge F_{2}
+\frac{1}{3}\chi^{3}F_{1}\wedge F_{1}
\Big),
\label{4dgv}
\end{align}
which is the standard Lagrangian describing 4D ${\cal N}=2$ supergravity coupled to a single vector multiplet with the prepotential
\be
P=\frac1{3\sqrt 3}\frac{X^3_1}{X_0}.
\ee
The symplectic coordinates $X_1,\,X_0$ are related to the physical scalars $\varphi,\,\chi$ via $X_1/X_0=\chi+i\sqrt3 e^{-\varphi/\sqrt3}$. 
The additional boundary piece in \eqref{eq:reduced-emcs-action} takes the form 
\begin{align}
I_{\rm B}
&= -\frac{1}{16\pi G_{5}}\frac{1}{3\sqrt{3}}
\int_{\mathcal{M}_{5}}
\mathrm{d}\left(3\star_{4}\mathrm{d}\varphi+
\chi^{2}A_{2}\wedge F_{1}
+2\chi A_{2}\wedge F_{2}
\right)\wedge\mathrm{d}z.
\label{5dbnd}
\end{align}
It should be noted that the above total derivative term is not globally defined when the gauge fields carry magnetic charge.

We now consider the dimensional reduction of the GHY boundary term \eqref{eq:5D-GHY}, 
\be
I_{\rm GHY}^{(5)}
= \frac{1}{8\pi G_{5}}
\int_{\partial\mathcal{M}_{5}}
\mathrm{d}^{4}x\,\sqrt{-\hat l}
\left(K[\hat l]-K[\hat{\bar{l}}]\right).
\label{eq:5D-GHY2}
\ee
From the reduction ansatz \eqref{eq:kk-ansatz}, one can deduce that 
\begin{align}
\sqrt{-\hat l}= e^{\frac{\varphi}{2\sqrt{3}}}\sqrt{-l},\quad \hat{r}_{\alpha} =e^{\frac{\varphi}{2\sqrt{3}}}r_{\alpha}, \quad \hat{r}^{\alpha}  =e^{-\frac{\varphi}{2\sqrt{3}}}r{}^{\alpha}.
\end{align}
where the hatted quantities are defined with respect to the 5D geometry, whereas unhatted quantities are defined using the 4D Einstein frame metric. 
Using $K[\hat l]=\hat\nabla_{\alpha}\hat r^{\alpha}$, we obtain
\begin{align}
\sqrt{-\hat l}K[\hat l]-\sqrt{-l}K[l]
&= \frac{1}{2\sqrt{3}}
\sqrt{-l}\,r^{\alpha}\partial_{\alpha}\varphi.
\label{kreduc}
\end{align}
Similarly, one can also carry out reduction of the reference background which has a constant background scalar $\bar{\varphi}$. Therefore the background subtraction terms are equal before and after the dimensional reduction. 

Collecting bulk and boundary terms from the dimensional reduction, we can further  simplify the results by noticing that the $\mathrm{d}\star_4\mathrm{d}\varphi$ in \eqref{5dbnd} can be directly integrated giving rise to two boundary terms. The one at the spatial infinity cancels precisely with the $\sqrt{-l}\,r^{\alpha}\partial_{\alpha}\varphi$ term from the reduction of the GHY term. The other one at the black string horizon simply vanishes.

After integrating over the $z$ direction and denoting $G_{4}=G_{5}/(2\pi R)$, we obtain the equality
\bea
\label{eq:action-reduction}
&&I_{\rm EMCS}^{(5)}+I_{\rm GHY}^{(5)}-I_{\rm B}= I^{(4)}_{2\partial}+I_{\rm GHY}^{(4)},\nn\\
&&I_{\rm B}= -\frac{1}{16\pi G_{4}}\frac{1}{3\sqrt{3}}
\int_{\mathcal{M}_{4}}
\mathrm{d}\Big(
\chi^{2}A_{2}\wedge F_{1}
+2\chi A_{2}\wedge F_{2}
\Big),\nn\\
\label{eq:4DAction}
&& I^{(4)}_{2\partial}
= \frac{1}{16\pi G_{4}}
\int_{\mathcal{M}_{4}}\mathcal{L}^{(4)},\quad 
I_{\rm GHY}^{(4)}=\frac{1}{8\pi G_{4}}
\int_{\partial\mathcal{M}_{4}}
\mathrm{d}^{3}x\,\sqrt{-l}
\left(K[l]-K[\bar{l}]\right),
\eea
where the right hand side is fully gauge invariant and captures all the solutions of the original 5D theory with the Killing direction $\partial_z$.

\subsection{Further truncation to 4D ${\cal N}=2$ minimal supergravity}
The action \eqref{4dgv} describes 4D supergravity coupled to a single vector multiplet. To truncate the theory further down to 4D $\mathcal N=2$ pure supergravity, we set both scalars to zero
\begin{equation}
\varphi=\chi=0.
\label{trunc}
\end{equation}
The $\varphi$ and $\chi$ field equations then imply
\begin{equation}
 3\star F_1\land F_1+\star F_2\land F_2=0,\quad - F_1\land\star F_2+\frac1{\sqrt 3}F_2\land F_2=0.
\end{equation}
The equations above are satisfied by identifying
\begin{equation}
 F_2=\sqrt3 \star F_1.
 \label{eq:F2F1}
\end{equation}
It can be readily shown that the remaining Einstein equation is derived from the action of 4D minimal supergravity
\begin{equation}
   \tilde{I}^{(4)}_{2\partial}=\frac{1}{16\pi G_4}\int_{\mathcal{M}_4} \mathrm{d}^4x\sqrt{-g}(R-F_1^2),
   \label{eq:einstein-maxwell-action}
\end{equation}
where $F_1$ is identified as the graviphoton. Comparing \eqref{trunc} to the general solution \eqref{eq:black-string-solution}, one sees that the truncation \eqref{trunc} can be achieved by choosing $\gamma=\delta=0$. The resulting solution is the uplift of the 4D electrically charged Kerr-Newman black hole. The details of the uplift are presented in App.~\ref{app:BS in KN}.


\section{Quantum statistical relation from Hamiltonian analysis}\label{sec:thermodynamic-hamiltonian}
In this section, we perform an ADM decomposition of the dimensionally reduced 4D theory \eqref{eq:4DAction} to recast it into Hamiltonian form~\cite{Arnowitt:1962hi,Regge:1974zd}. We then use the resulting Hamiltonian to compute the thermodynamic quantities explicitly.

We now turn to the Hamiltonian formulation of the theory, which provides a complementary perspective on these thermodynamic relations. By keeping the boundary terms explicit, we can identify the individual thermodynamic contributions and establish their connection with the on-shell action. For stationary solutions, the on-shell action $I_{\text{on-shell}}$ is equal to inverse temperature $\beta$ multiplying the on-shell Hamiltonian $H_{\text {on-shell}}$ \footnote{Here, we also use $\beta$ to denote the inverse temperature in our thermodynamic analysis, its meaning should be clear from the context and should not be confused with the parameter $\beta$ appearing in the solution. }. Explicit computation shows that $H_{\text{ on-shell}}$ is precisely equal to the free energy, therefore establishing the quantum statistical relation.

\subsection{ADM Formalism}
We now express $I^{(4)}_{2\partial}+I_{\rm GHY}^{(4)}$ in Hamiltonian form~\cite{Hawking:1995fd}. First, we introduce a future-directed timelike unit normal $n_\mu$ and spacelike unit normals $r_\mu$ and $m_\mu$ pointing toward increasing $r$ and $x$, respectively,
\begin{equation}
    n_{\mu}=(-N,0,0,0),\quad
    r_{\mu}=(0,1/\sqrt{g^{rr}},0,0),\quad
    m_{\mu}=(0,0,1/\sqrt{g^{xx}},0),
\end{equation}
where $N$ is the lapse function and the coordinates are ordered as $(t,r,x,\phi)$. The normal vectors therefore satisfy
\begin{equation}
    n^{\mu}n_{\mu}=-1,\quad
    r^{\mu}r_{\mu}=1,\quad
    m^{\mu}m_{\mu}=1,\quad
    n^{\mu}r_{\mu}=n^{\mu}m_{\mu}=r^{\mu}m_{\mu}=0.
\end{equation}
We define the corresponding induced metrics
\begin{align}
    h_{\mu\nu}&=g_{\mu\nu}+n_{\mu}n_{\nu},
    \qquad
    l_{\mu\nu}=g_{\mu\nu}-r_{\mu}r_{\nu},
\end{align}
and
\begin{align}
    \sigma_{\mu\nu}=g_{\mu\nu}+n_{\mu}n_{\nu}-r_{\mu}r_{\nu},\qquad
    \theta_{\mu\nu}=g_{\mu\nu}+n_{\mu}n_{\nu}-m_{\mu}m_{\nu}.
\end{align}
The scalar-field gradients and gauge fields are decomposed as
\bea
    \nabla_{\mu}\varphi
    &=&-n_{\mu}\dot{\varphi}+\mathrm{D}_{\mu}\varphi,
    \qquad
    \nabla_{\mu}\chi
    =-n_{\mu}\dot{\chi}+\mathrm{D}_{\mu}\chi,
    \nn\\
    A_{1,\mu}
    &=&-A_{1,\boldsymbol{n}}n_{\mu}
    +\tilde{A}_{1,\mu},\quad  A_{1,\boldsymbol{n}}
    =n^{\mu}A_{1,\mu},
    \nn\\
    A_{2,\mu}
    &=&-A_{2,\boldsymbol{n}}n_{\mu}
    +\tilde{A}_{2,\mu},\quad A_{2,\boldsymbol{n}}
    =n^{\mu}A_{2,\mu}.
\eea
In the expression above, the overdot denotes the Lie derivative along $n^\mu\partial_\mu$, $\mathrm{D}_{\mu}$ is the covariant derivative with respect to $h_{\mu\nu}$ and the tilde denotes the projection to spatial directions orthogonal to $n^\mu$. For gauge field strength we have
\begin{align}
    F_{1,\mu\nu}
    &=2\Big(
    \dot{\tilde{A}}_{1,[\mu}n_{\nu]}
    +A_{1,\boldsymbol{n}}n_{[\mu}a_{\nu]}
    +\mathrm{D}_{[\mu}\tilde{A}_{1,\nu]}   +n_{[\mu}\mathrm{D}_{\nu]}A_{1,\boldsymbol{n}}
    \Big),\\
    F_{2,\mu\nu}
    &=2\Big(
    \dot{\tilde{A}}_{2,[\mu}n_{\nu]}
    +A_{2,\boldsymbol{n}}n_{[\mu}a_{\nu]}
    +\mathrm{D}_{[\mu}\tilde{A}_{2,\nu]}    +n_{[\mu}\mathrm{D}_{\nu]}A_{2,\boldsymbol{n}}
    \Big),
\end{align}
with $a_\mu=n^\nu\nabla_\nu n_\mu$ denoting the normal acceleration.

Substituting ADM decomposition of fields into the 4D Lagrangian, performing integration by parts, we obtain a Lagrangian density $\mathcal{L}_{\rm ADM}$ involving only the first order derivatives of the fields.
The canonical momenta of the metric and scalar fields are obtained in the usual way\footnote{For convenience, we omit the overall factor $1/(16\pi G_4)$ in the definition of the conjugate momenta.}
\begin{align}
\pi^{ij} & \coloneqq\frac{\partial\mathcal{L}_{\rm ADM}}{\partial(\partial_{t}h_{ij})}=\sqrt{h}(K^{ij}-Kh^{ij}),\\
\pi_{\varphi} & \coloneqq\frac{\partial\mathcal{L}_{\rm ADM}}{\partial(\partial_{t}\varphi)}=\sqrt{h}\dot{\varphi},\\
\pi_{\chi} & \coloneqq\frac{\partial\mathcal{L}_{ADM}}{\partial(\partial_{t}\chi)}=\sqrt{h}e^{\frac{2\varphi}{\sqrt{3}}}\dot{\chi},
\end{align}
where
\begin{equation}
    K_{ij}=\frac{1}{2N}    (\partial_{t}h_{ij}-2\mathrm{D}_{(i}N_{j)}).
\end{equation}
The canonical momenta of gauge-fields are
\begin{align}
\pi_1^{\mu}\coloneqq\frac{\partial\mathcal{L}_{\rm ADM}}{\partial(\partial_{t}\tilde{A}{}_{1,\mu})} & =\sqrt{h}\Big(-e^{-\sqrt{3}\varphi}F{}_{1}^{\mu\nu}n_{\nu}-e^{-\frac{\varphi}{\sqrt{3}}}\chi^{2}F{}_{1}^{\mu\nu}n_{\nu}-e^{-\frac{\varphi}{\sqrt{3}}}\chi F{}_{2}^{\mu\nu}n_{\nu}\nonumber \\
 & \quad\quad +\frac{1}{3\sqrt{3}}\epsilon^{\mu\nu\rho}\chi^{3}F_{1,\nu\rho}+\frac{1}{2\sqrt{3}}\epsilon^{\mu\nu\rho}\chi^{2}F_{2,\nu\rho}\Big),
 \\
\pi_{2}^{\mu}\coloneqq\frac{\partial\mathcal{L}_{\rm ADM}}{\partial(\partial_{t}\tilde{A}{}_{2,\mu})} & =\sqrt{h}\Big(-e^{-\frac{\varphi}{\sqrt{3}}}\chi F{}_{1}^{\mu\nu}n_{\nu}-e^{-\frac{\varphi}{\sqrt{3}}}F{}_{2}^{\mu\nu}n_{\nu}
\nonumber \\
 & \quad\quad +\frac{1}{2\sqrt{3}}\epsilon^{\mu\nu\rho}\chi^{2}F_{1,\nu\rho}+\frac{1}{\sqrt{3}}\epsilon^{\mu\nu\rho}\chi F_{2,\nu\rho}\Big),
\end{align}
where the spatial Levi-Civita tensor is the contraction of $n^\mu$ with the 4D Levi-Civita tensor $\epsilon^{\mu\nu\rho}
=n_{\sigma}\epsilon^{\sigma\mu\nu\rho}$.

From the Legendre transformation of the Lagrangian, we obtain the total Hamiltonian as 
\begin{align}\label{eq:adm-hamiltonian}
    H&=\frac{1}{16\pi G_{4}}\int_{\Sigma_{t}}\mathcal{H}_{\rm C}\,\mathrm{d}^{3}x+H_{B},
\end{align}
where $\Sigma_{t}$ denotes the constant time slice. The bulk Hamiltonian density ${\cal H}_C$ is a linear combination of the constraints and thus vanishes on the constraint surface. Its explicit form is given in App. \ref{app:ADM-constraints}.
Here $H_{B}$ denotes the boundary Hamiltonian, which contains both the gravitational and gauge-field surface contributions. 

We next specify the boundary Hamiltonian $ H_B$, which receives contributions from the spatial infinity and the horizon in general. Moreover, when the gauge fields carry magnetic charges, such as the black string solution, their proper treatment requires two coordinate patches. In our case, we choose the two patches to be northern and southern hemispheres overlapping on the equator at $x=0$. Denoting  equatorial boundaries of northern and  southern hemispheres by $\Sigma_+$ and $\Sigma_{-}$, corresponding to $x=0^{+}$ and $x=0^{-}$ respectively, one can recast the boundary Hamiltonian in the form
\begin{align}
 \label{eq:boundary-hamiltonian}
H_B& =-\frac{1}{16\pi G_{4}}\int_{\Sigma_{\infty}}2N(k-k_{0})\sqrt{\sigma}\,\text{d}x\,\text{d}\phi-\frac{1}{16\pi G_{4}}\int_{\Sigma_{{\text{h}}}}2N(r_{\mu}n^{\nu}\nabla_{\nu}n^{\mu})\sqrt{\sigma}\,\text{d}x\,\text{d}\phi\nonumber \\
& \ \ +\frac{1}{16\pi G_{4}}\int_{\Sigma_{\infty}}\mathcal{B}_{\mu}r^{\mu}\sqrt{\sigma}\,\text{d}x\,\text{d}\phi-\frac{1}{16\pi G_{4}}\int_{\Sigma_{\text{h}}}\mathcal{B}_{\mu}r^{\mu}\sqrt{\sigma}\,\text{d}x\,\text{d}\phi\nonumber \\
 & \ \ -\frac{1}{16\pi G_{4}}\int_{\Sigma_{+}}\mathcal{B}_{\mu}m^{\mu}\sqrt{\theta}\,\text{d}r\,\text{d}\phi+\frac{1}{16\pi G_{4}}\int_{\Sigma_{-}}\mathcal{B}_{\mu}m^{\mu}\sqrt{\theta}\,\text{d}r\,\text{d}\phi,
\end{align}
where $k$ is the trace of the extrinsic curvature of the two-dimensional surface at $r=\infty$ and $t={\rm const}$, and $k_{0}$ is its value embedded in flat background~\cite{Brown:1992br}. $\Sigma_{\rm h}$ denotes the constant time slice of the horizon and we have defined 
\begin{equation}
\mathcal{B}_{\mu}=\frac{1}{\sqrt{h}}\Big(2N^{\nu}\pi_{\mu\nu}+A_{1,t}\pi_{1,\mu}+A_{2,t}\pi_{2,\mu}\Big),\quad A_{1,t}
    \equiv NA_{1,\boldsymbol{n}}
    +N^{i}\tilde{A}_{1,i}, \quad A_{2,t}
    \equiv NA_{2,\boldsymbol{n}}
    +N^{i}\tilde{A}_{2,i}.
\end{equation}

\subsection{Thermodynamics}
Upon dimensional reduction, the 5D black string gives rise to an asymptotically flat 4D black hole carrying electric charge under the gauge field $A_{1,\mu}$ and both electric and magnetic charges under $A_{2,\mu}$.
For a magnetically charged black string, the gauge field $A_{2,\mu}$ must be defined on two patches covering the northern and southern hemispheres overlapping on the equator at $x=0$~\cite{Copsey:2005se,Hanaki:2007mb}. Therefore in the evaluation of the Hamiltonian, in the northern patch, $x$ ranges over $(0,1]$ while in the southern patch, $x$ ranges over $[-1,0)$.

Following \cite{Brown:1990fk}, we shall evaluate the Hamiltonian in a coordinate system corotating with the horizon. To do so, we change the coordinates through
\begin{equation}
   t=t^{\prime},\quad r=r',\quad x=x',\quad
   \phi   =\phi^{\prime}+\Omega_{\phi}^{(4)}t^{\prime},
    \quad
     \Omega_{\phi}^{(4)}=-\frac{g_{t\phi}}{g_{\phi\phi}}\bigg|_{r=r_{\text{h}}}.
\end{equation}
In the new coordinate system, 
the ADM variables satisfy the following conditions on the horizon and near spatial infinity 
\begin{align}
    N^{\prime}\big|_{r=r_{\rm h}}
    &=0,\qquad
    N^{\prime i}\big|_{r=r_{\rm h}}=(0,0,0),\\
    N^{\prime}\big|_{r\rightarrow\infty}
    &=1,\qquad
    N^{\prime i}\big|_{r\rightarrow\infty}
    =(0,0,\Omega_{\phi}^{(4)}),
\end{align}
where the spatial components are ordered as $(r',x',\phi^{\prime})$. Adding a constant shift  to the temporal component of the gauge fields, the asymptotic gauge potentials take the form
\begin{align}
    A_{1,\mu}^{\prime}\big|_{r\rightarrow\infty}
    &=(c_{1,t},0,0,c_{1,\phi}),\label{eq:asy-A1}\\
    A_{2,\mu}^{\prime}\big|_{r\rightarrow\infty}
    &=\Big(
    c_{2,t}-\Omega_{\phi}^{(4)}\Lambda x,
    0,0,-\Lambda x+c_{2,\phi}
    \Big).
    \label{eq:asy-A2}
\end{align}
The Killing vector null on the horizon now coincides with $\partial_t'$ as
\begin{align}
    \xi
&=\partial_{t}+\Omega_{\phi}^{(4)}\partial_{\phi}=\Big(
    \partial_{t^{\prime}}
    -\Omega_{\phi}^{(4)}\partial_{\phi^{\prime}}
    \Big)
    +\Omega_{\phi}^{(4)}\partial_{\phi^{\prime}}
    =\partial_{t^{\prime}}.
\end{align}

Below we choose $c_{n,t},\,(n=1,2)$ in such a way that the electric potential is continuous across the equator. We could also impose $A_\mu A^\mu|_{r=r_h}=0$ on the horizon by choosing $c_{n,\phi},\,(n=1,2)$. However, this extra condition is never needed in the computation of the on-shell Hamiltonian which is finite as long as $A_t$ is finite. Thus, the situation in the gauge invariant 4D formulation is different from that of the original 5D description, where the on-shell action depends on the value of $A_t|_{r=r_h}$.

For notational simplicity, we will omit the prime in the superscript throughout the remainder of this subsection. The bulk Hamiltonian vanishes on the solution. Thus the total Hamiltonian is given by its boundary value. Also because the temporal component of the gauge fields and the conjugate momenta are continuous across the equator, the two terms in \eqref{eq:boundary-hamiltonian} associated with $\Sigma_\pm$ cancel. We are then left with
\begin{align}
H &=-\frac{1}{8\pi G_4}\int_{\Sigma_{\infty}}N(k-k_{0})\sqrt{\sigma}\,\mathrm{d}x\,\mathrm{d}\phi-\frac{\Omega_{\phi}^{(4)}}{16\pi G_4}
\int_{\Sigma_{\infty}}\sqrt{\sigma}
\big(-2K_{\mu\nu}r^{\nu}\bar\phi^{\mu}\big)\text{d}x\,\text{d}\phi
\nonumber\\
&\ \ -\frac{1}{16\pi G_4}\int_{\Sigma_{\infty}}
\Big(A_{1,t}\rho_{1}
+A_{2,t}\rho_{2}\Big)
\sqrt{\sigma}\,\mathrm{d}x\,\mathrm{d}\phi+\frac{1}{16\pi G_4}\int_{\Sigma_{\text{h}}}
\Big(A_{1,t}\rho_{1}
+A_{2,t}\rho_{2}\Big)
\sqrt{\sigma}\,\mathrm{d}x\,\mathrm{d}\phi
\nn\\
&\quad -\frac{1}{16\pi G_4}\int_{\Sigma_{\text{h}}}2N(r_{\mu}n^{\nu}\nabla_{\nu}n^{\mu})\sqrt{\sigma}\,\mathrm{d}x\,\mathrm{d}\phi,
\end{align}
in which $\bar\phi^{\mu}=(0,0,0,1)$ and the electric charge density is defined as $\rho_{n}:=-\pi_{n,\mu}r^{\mu}/\sqrt{h}$. The meaning of each term in the expression above is evident. The first term gives the ADM mass $M^{(4)}$. The second term is $-\Omega^{(4)}_{\phi}J^{(4)}_{\phi}$. The third and the fourth terms combine into $-\Phi^{(4)}_{\rm E,1}\,Q^{(4)}_{\rm E,1}-\Phi^{(4)}_{\rm E,2}\,Q^{(4)}_{\rm E,2}$ and the last term evaluates to $-T^{(4)}S^{(4)}$, as $N(r_{\mu}n^{\nu}\nabla_{\nu}n^{\mu})$ on the horizon is equal to the surface gravity. To compute the third and the fourth terms, we first notice that the Gauss law ensures that the electric charge measured at the horizon is equal to that measured at infinity. There is some subtlety regarding the definition of the electric potential which we explain as follows. The standard prescription of electric potential is given by the difference of the quantity $\xi^{\mu}A_{\mu}$ between spatial infinity and horizon. In the corotating frame, where $\xi=\partial_{t}$, $\xi^{\mu}A_{\mu}=A_t$. For the general black string solution, both $A_{1,t}$ and $A_{2,t}$ are constant on the horizon. While $A_{1,t}$ approaches constant values near spatial infinity, $A_{2,t}$ behaves as
\be
A_{2,t}|_{r\rightarrow\infty}\rightarrow c_{2,t}-\Omega^{(4)}_{\phi}\Lambda\,x,
\ee
where there is a dipole contribution proportional to $x$. Because the charge density $\rho_{2}$ is an even function of $x$, the dipole piece does not contribute to the boundary Hamiltonian. We therefore define the electric potential as 
\begin{equation}
    \Phi_{{\rm E},1}^{(4)}=A_{1,t}\big|_{r_{\rm h}}^{\infty},\quad \Phi_{{\rm E},2}^{(4)}=c_{2,t}-A_{2,t}|_{r=r_{\rm h}}
    .
    \label{eq:def-potential}
\end{equation}
Evaluating each term in the boundary Hamiltonian explicitly and using
$G_{5}=2\pi R\,G_{4}$, we find the following agreement between 4D and 5D  conserved charges 
\begin{align}
    &M^{(4)}=M^{(5)},\quad
    S^{(4)}=S^{(5)},\quad
    J_{\phi}^{(4)}=J_{\phi}^{(5)},\quad Q_{{\rm E},1}^{(4)}=P_{z}^{(5)},\quad
    Q_{{\rm E},2}^{(4)}=Q_{\rm E}^{(5)},
    \label{eq:4d-5d-charges}
\end{align}
and similarly for thermodynamic potentials 
\begin{align}
&T^{(4)}=T^{(5)},\quad\Omega_{\phi}^{(4)}=\Omega_{\phi}^{(5)},\quad \Phi_{{\rm E},1}^{(4)}=v_{z}^{(5)},\quad\Phi_{{\rm E},2}^{(4)}=\Phi_{\rm E}^{(5)}.
    \label{eq:4d-5d-potentials}
\end{align}
The thermodynamic correspondence also extends to the magnetic sector. The magnetic charge associated with $A_2$ and its conjugate potential satisfy
\begin{align}
    Q_{{\mathrm M},2}^{(4)}&=Q_{\mathrm M}^{(5)},
    \qquad
    \Phi_{{\mathrm M},2}^{(4)}=\Phi_{\mathrm M}^{(5)}.
\end{align}
The 5D quantities are given in Eqs.~\eqref{eq:5D-thermodynamics-a} and \eqref{eq:5D-thermodynamics-b}, with the extensive quantities integrated over the string direction. These identifications establish the thermodynamic correspondence between the reduced black hole and the black string within the Hamiltonian formulation. In particular, the mass and electric-charge identifications provide a Hamiltonian realization of the dimensional-reduction relations discussed in Eqs.~(2.36)--(2.38) of Ref.~\cite{Compere:2010fm}, with the charge normalizations and sign conventions translated to ours. Finally, using the relation between Euclidean on-shell action and the total on-shell Hamiltonian, we immediately obtain
the quantum statistical relation
\begin{equation}\label{eq:4d-gibbs}
  I_{\text{on-shell}}^{(4)}=\beta(M^{(4)}-T^{(4)}S^{(4)}-\Omega_{\phi}^{(4)} J_{\phi}^{(4)}-\Phi_{{\rm E},1}^{(4)}Q_{{\rm E},1}^{(4)}-\Phi_{{\rm E},2}^{(4)}Q_{{\rm E},2}^{(4)}).
\end{equation}
To our knowledge, this is the first explicit derivation of the quantum statistical relation between a gauge-invariant reduced on-shell action and the Gibbs free energy for this black string family, obtained directly from the ADM boundary Hamiltonian.

As discussed in Sec.~\ref{sec:supergravity-setup}, the 5D action acquires gauge-dependent surface contributions from the Chern--Simons coupling. In fact these surface terms are generically nonzero on the black string solutions, thus relating  $I_{\text{on-shell}}^{(5)}$ to the Gibbs free energy $G_{\text{free}}^{(5)}$ requires an additional gauge-compensating term. The 4D action constructed here is gauge invariant. Together with the charge correspondence established above, our Hamiltonian derivation shows that this reduced action directly reproduces the Gibbs free energy through Eq.~\eqref{eq:4d-gibbs}, without further gauge compensation. This provides a gauge-invariant action formulation of the thermodynamics of the black string family via dimensional reduction.


\section{Leading higher derivative corrections}\label{sec:higher-derivative-corrections}
We now move on to test the validity of our approach by considering leading higher derivative corrections to the black string on-shell action. In 5D minimal supergravity, the leading higher derivative corrections are given in terms of supersymmetric completion of Weyl tensor squared \cite{Hanaki:2006pj, Ozkan:2013uk,Ozkan:2013nwa}. To illustrate the main idea while  evading complications caused by higher  derivative terms, we specialize to $\gamma=\delta=0$ case.  
As shown in Sec.~\ref{sec:KK-reduction}, the $\gamma=\delta=0$ black string solution can be dimensionally reduced to 4D Kerr-Newman black hole described by 4D $\mathcal N=2$ minimal supergravity. Thus, we expect that in the presence of 5D supersymmetric Weyl tensor squared action, the on-shell action obtained via dimensional reduction should agree with previous results \cite{Hu:2025ogz} obtained for 4D Kerr-Newman black hole with curvature squared corrections. 
We emphasize that similar to the 2-derivative case, the total derivative terms arising from the reduction are crucial for relating the 5D and 4D on-shell actions.

 The supersymmetric 4-derivative invariants consist of three parts \cite{Hanaki:2006pj,Ozkan:2013uk,Ozkan:2013nwa,Gold:2023ymc,Ozkan:2024euj}: Weyl tensor squared action
$\mathcal{L}_{{\rm Weyl}^{2}}$, Ricci tensor squared action
$\mathcal{L}_{{\rm Ricci}^{2}}$ and Ricci scalar squared action $\mathcal{L}_{R^{2}}$. After some  field redefinitions and keeping terms up to first order in 4-derivative coupling constants, the action gives the supersymmetric Einstein-Weyl (EW) action \cite{Ma:2024ynp,Hu:2025ogz}
\begin{align}
I_{\rm EW}
&=\frac{1}{16\pi G_{5}}
\int_{\mathcal{M}_5}\sqrt{-\hat{g}}\,\mathrm{d}^{5}x
(\mathcal{L}_{\rm EMCS}^{(5)}+c_{1}\mathcal{L}_{4\partial,{\rm EW}}^{(5)}),
\nonumber\\
\mathcal{L}_{4\partial,{\rm EW}}^{(5)}
&=\hat{C}_{\alpha\beta\gamma\delta}\hat{C}^{\alpha\beta\gamma\delta}
-\frac{1}{2}\hat{C}_{\alpha\beta\gamma\delta}
\hat{F}^{\alpha\beta}\hat{F}^{\gamma\delta}
+\frac{13}{96}(\hat{F}^{\alpha\beta}\hat{F}_{\alpha\beta})^{2}
\nonumber\\
&\quad
-\frac{13}{24}\hat{F}_{\alpha\beta}\hat{F}^{\beta\gamma}
\hat{F}_{\gamma\pi}\hat{F}^{\pi\alpha}
-\frac{\hat{\epsilon}^{\alpha\beta\gamma\delta\zeta}}{2\sqrt{3}}
\hat{A}_{\alpha}
\hat{C}_{\beta\gamma}^{\ \ \ \eta\kappa}
\hat{C}_{\delta\zeta\eta\kappa}.
\label{eq:einstein-weyl-action}
\end{align}
where $\mathcal{L}_{\rm EMCS}^{(5)}$ is the EMCS theory \eqref{eq:5DEMCS}. With the additional redefinition of the metric
\begin{equation}
\hat{g}_{\alpha\beta}\rightarrow\hat{g}_{\alpha\beta}+c_1\big(-\frac{4}{3}\hat{R}_{\alpha\beta}+\frac{1}{3}\hat{R}\hat{g}_{\alpha\beta}-\frac{1}{36}\hat{g}_{\alpha\beta}\hat{F}_{\gamma\delta}\hat{F}^{\gamma\delta}\big),
\end{equation}
and keeping terms up to first order in $c_1$, the Lagrangian $\mathcal{L}_{4\partial,{\rm EW}}^{(5)}$ can be further simplified as
\begin{align}
\mathcal{L}_{4\partial}^{(5)}
={}&\hat{R}_{\alpha\beta\gamma\delta}
\hat{R}^{\alpha\beta\gamma\delta}
-\frac{1}{2}\hat{R}_{\alpha\beta\gamma\delta}
\hat{F}^{\alpha\beta}\hat{F}^{\gamma\delta}
+\frac{5}{36}(\hat{F}^{2})^{2}
-\frac{13}{24}{\rm Tr}\hat{F}^{4}\nonumber\\
&-\frac{1}{2\sqrt{3}}\hat{\epsilon}^{\alpha\beta\gamma\delta\xi}
\hat{A}_{\alpha}
\hat{R}_{\beta\gamma}{}^{\eta\kappa}
\hat{R}_{\delta\xi\eta\kappa}.
\label{eq:hd-5d-riemann}
\end{align}
We then apply Reall-Santos method \cite{Reall:2019sah,Hu:2023gru} to evaluate 4-derivative corrections to the on-shell action.
This amounts to plugging in the uncorrected black string below into \eqref{eq:hd-5d-riemann} and performing the integration. The 5D black string solution with $\gamma=\delta=0$ corresponds to the uplift of 4D Kerr-Newman black hole (see App.~\ref{app:BS in KN} for details)
\begin{align}
\mathrm{d}s^{2}_{5}&=\mathrm{d}s^{2}_{4}+(\mathrm{d}z+A_{1})^{2}, &
\hat{A}&=A_{2}, & \varphi&=\chi=0,\nonumber\\
\hat{e}^{a}&=e^{a}, & \hat{e}^{z}&=\mathrm{d}z+A_{1}, &
F_{2}&=\sqrt{3}\star_4 F_{1}.
\label{eq:hd-reduction-ansatz}
\end{align}
Before carrying out the integration, we can first  manipulate the integrand using the algebraic relation among different fields implied by the solution \eqref{eq:hd-reduction-ansatz}.
 After a long calculation, the 5D integrand can be brought into a fully gauge invariant form, by subtracting  a non-gauge invariant total derivative term,  
\begin{equation}
\mathcal{L}_{4\partial}^{(5)}\star_{5}\boldsymbol{1}-\mathcal{L}^{{\rm TD}}_{4\partial}=  \mathcal{L}_{4\partial}^{(4)}\star_{4}\boldsymbol{1}\wedge \mathrm{d}z.
\label{eq:hd-reduction-identity}
\end{equation}
in which
\begin{align}
\mathcal{L}_{4\partial}^{(4)}= & R_{abcd}R^{abcd}+\frac{3}{2}RF^{2}_{1}-8R_{ab}F^{ac}_{1}F_{1}{}^{b}{}_{c}+2\nabla_{a}F^{ac}_{1}\nabla_{b}F_{1}{}^{b}{}_{c}, \label{eq:hd-4d-LeadingAction}\\ 
\mathcal{L}^{{\rm TD}}_{4\partial}= & \nabla_{c}V^{c}\star_{4}\boldsymbol{1}\wedge \mathrm{d}z+\frac{1}{\sqrt{3}}\mathrm{d}\big(F_{1,ab}F_{2}\wedge\omega^{ab}\wedge \mathrm{d}z+2A_{2}\wedge\widehat{{\rm CS}}\big),
\label{eq:hd-4d-density}
\end{align}
and
\begin{equation}
    \mathrm{d}\widehat{{\rm CS}}={\rm Tr}(\hat{R}\wedge\hat{R}),\  \ \
\widehat{{\rm CS}}=\hat{\omega}^{A}_{\ B}\land\hat{R}^{B}_{\ A}-\frac{1}{3}\hat{\omega}^{A}_{\ B}\land\hat{\omega}^{B}_{\ C}\land\hat{\omega}^{C}_{\ A},
\end{equation}
where the uppercase Latin letters $A,B,C,\ldots$ denote five-dimensional tangent-space (Lorentz) indices. We have defined $F_1^2=F_{1,ab}F_1^{ab}$ and $V^{c}={}F_{1,ab}\nabla^{c}F_{1}^{ab}-2F_{1}^{ac}\nabla^{b}F_{1,ab}$. Readers who are interested in the details can refer to 
App.~\ref{app:mixed-chern-simons-reduction}.

The right hand side of \eqref{eq:hd-reduction-identity} can be easily evaluated by first noticing that it is on-shell equivalent to Gauss-Bonnet invariant, using the fact that the leading order solution obeys 
\be
R=0,\quad \nabla_a F^{ab}_1=0,\quad R_{ab}F^{ac}_{1}F_{1}{}^{b}{}_{c}=\frac12 R_{ab}R^{ab}.
\ee
For the smooth Euclidean Kerr--Newman solution, the  integral of the Gauss--Bonnet invariant evaluates to $64\pi^2$ \cite{Ma:2003uj,Charles:2016wjs}.

We also verified the equality \eqref{eq:hd-4d-density} by performing the integration on the leading order solution explicitly. We find that 
the 5D bulk action \eqref{eq:hd-5d-riemann} and the total derivative term yield
\bea
I_{4\partial}^{(5)}
&=&-\frac{2\pi^2 R \big[q^{2}(q^{2}-r_{0}^{2}-a^{2})
+4r_{0}^{2}(r_{0}^{2}-q^{2}-a^{2})\big]}
{G_{5}r_{0}^{2}(r_{0}^{2}-a^{2}-q^{2})},
\nn\\
 I_{4\partial}^{\rm TD }&=&\frac{2\pi^2  q^{2}R(a^{2}+r_{0}^{2}-q^{2})}
{G_{5}r_{0}^{2}(r_{0}^{2}-a^{2}-q^{2})}.
\label{eq:total-derivative-action}
\eea
The integral of the first term in \eqref{eq:hd-4d-density} vanishes on the solution, while to compute the total derivative term, we adopt the vielbein
\begin{align}
\mathrm{d}s^{2}_{5} & =\hat{\eta}_{AB}\hat{e}^{A}\hat{e}^{B}=\eta_{ab}e^a e^b+(\hat{e}^{z})^{2},\nonumber \\
\hat{e}^{0} & =\sqrt{\frac{\Delta}{\rho}}\big(\mathrm{d}t-a(1-x^{2})\mathrm{d}\phi\big),\ \ \hat{e}^{1}=\sqrt{\frac{\rho}{\Delta}}\mathrm{d}r,\ \ \hat{e}^{2}=\sqrt{\frac{\rho}{1-x^{2}}}\mathrm{d}x,\nonumber \\
\hat{e}^{3} & =\sqrt{\frac{1-x^{2}}{\rho}}\big(a\mathrm{d}t-(r^{2}+a^{2})\mathrm{d}\phi\big),\ \ \ \hat{e}^{z}=\mathrm{d}z+A_{1}.
\label{eq:kn-vielbein}
\end{align}
Furthermore, the total derivative terms in~\eqref{eq:hd-4d-density} receive contributions from the boundary of northern and southern hemispheres, including the black string horizon, asymptotic infinity and equatorial interfaces.

Miraculously, the left hand side of \eqref{eq:hd-reduction-identity} gives rise to a pure constant
  \be
  I_{4\partial}^{(5)}-I_{4\partial}^{\rm TD }=-\frac{8\pi^2 R}{G_5}=-\frac{4\pi}{G_4}.
  \label{4derKN}
  \ee

Recall that at the leading order, the proper action for the dimensionally reduced black string with $\gamma=\delta=0$ is given by \eqref{eq:einstein-maxwell-action} which combining with \eqref{4derKN}, gives rise to the 4-derivative corrected on--shell action for black string with $\gamma=\delta=0$ 
\be
I_{2\partial+4\partial}=\frac{\beta^{(4)}}{2G_4}\big(\mu-\frac{q^2 r_0}{r_0^2+a^2} \big)-c_1\frac{4\pi}{G_4}.
\ee

This result is in perfect agreement with the on-shell action of 4D Kerr-Newman in $\mathcal N=2$ minimal supergravity extended by 4-derivative corrections. It is known that via field redefinitions, all the 4-derivative actions in 4D $\mathcal N=2$ supergravity can be reduced to Gauss-Bonnet invariant \cite{Hu:2023qhs}.


\section{Discussion}\label{sec:discussion}
In this work, we developed a gauge-invariant formulation of the thermodynamics of rotating black strings in five-dimensional minimal supergravity through Kaluza–Klein reduction. By carefully accounting for the boundary contributions induced by the Chern–Simons coupling, we obtained a gauge-invariant four-dimensional action and established the quantum statistical relation between its Euclidean on-shell value and the Gibbs free energy using the Hamiltonian formalism. We further investigated leading four-derivative corrections to the rotating magnetic black string whose dimensional reduction yields the Kerr–Newman black hole. After electromagnetic dualization, the reduced theory provides a suitable thermodynamic ensemble for evaluating these corrections. We showed that the four-derivative Lagrangian reduces to the Gauss–Bonnet density on the leading-order Kerr–Newman background, giving a constant entropy correction $4\pi c_1/G_4$. The total derivative terms arising from dimensional reduction precisely account for the difference between the 5D and 4D on-shell action corrections.

It would be interesting to extend our analysis to more general rotating black strings with nontrivial scalar profiles and to investigate their higher-derivative thermodynamics, including possible supersymmetric limits. A direct Hamiltonian analysis in five dimensions and a systematic treatment of Chern–Simons boundary contributions could provide further insights into the thermodynamics of higher-dimensional black objects.


\section*{Acknowledgements}
We are particularly grateful to Yi Pang for insightful discussions and valuable suggestions. We also thank Kiril Hristov, Liang Ma, and Robert Saskowski for helpful discussions. Y.Y. is supported by the National Natural Science Foundation of China under Grant No. 12547120. P.H. is supported by the Postdoctoral Fellowship Program of the China Postdoctoral Science Foundation under Grant No. GZC20252203 and the National Natural Science Foundation of China under Grant No. 12547116. Yi Pang's contribution to this work was supported by the National Natural Science Foundation of China under Grant No. 12575076.

\appendix

\section{Explicit expressions for the ADM constraints}
\label{app:ADM-constraints}
The bulk Hamiltonian density $\mathcal{H}_{\rm C}$ is a linear combination of the constraints
\begin{align}
    \mathcal{H}_{\rm C}
    &=NC+N^{i}C_{i}
    +\Big(NA_{1,\boldsymbol{n}}    +N^{i}\tilde{A}_{1,i}\Big)\mathcal{C}^{(1)}
    +\Big(NA_{2,\boldsymbol{n}}   +N^{i}\tilde{A}_{2,i}\Big)\mathcal{C}^{(2)},
\end{align}
where the Hamiltonian constraint $C$ is 
\begin{align}
C & =\frac{1}{\sqrt{h}}\pi{}_{ij}\pi{}^{ij}-\frac{1}{2\sqrt{h}}\pi{}^{i}{}_{i}\pi{}^{j}{}_{j}-\sqrt{h}{}^{3}\!R\nonumber \\
 & \ \ +\frac{1}{2\sqrt{h}}\pi_{\varphi}^{2}+\frac{1}{2\sqrt{h}}e^{-\frac{2\varphi}{\sqrt{3}}}\pi_{\chi}^{2}+\frac{\sqrt{h}}{2}\text{D}_{i}\varphi\text{D}^{i}\varphi+\frac{\sqrt{h}}{2}e^{\frac{2\varphi}{\sqrt{3}}}\text{D}_{i}\chi\text{D}^{i}\chi\nonumber \\
 & \ \ +\frac{1}{2\sqrt{h}}e^{\sqrt{3}\varphi}\pi_{1,i}\pi_1^{i}-\frac{1}{\sqrt{h}}e^{\sqrt{3}\varphi}\chi\pi_1^{i}\pi_{2,i}+\frac{1}{2\sqrt{h}}\Big(e^{\frac{\varphi}{\sqrt{3}}}+e^{\sqrt{3}\varphi}\chi^{2}\Big)\pi_{2,i}\pi_2^{i}\nonumber \\
 &\ \ +\frac{\sqrt{h}}{108}e^{-\sqrt{3}\varphi}(3+e^{\frac{2\varphi}{\sqrt{3}}}\chi^{2})\bigg[(3+e^{\frac{2\varphi}{\sqrt{3}}}\chi^{2})^{2}\tilde{F}_{1,ij}\tilde{F}{}_{1}^{ij}\nonumber \\
 &\ \ \ \ +3e^{\frac{2\varphi}{\sqrt{3}}}\Big(2\chi(3+e^{\frac{2\varphi}{\sqrt{3}}}\chi^{2})\tilde{F}_{1,ij}\tilde{F}{}_{2}^{ij}+3(1+e^{\frac{2\varphi}{\sqrt{3}}}\chi^{2})\tilde{F}_{2,ij}\tilde{F}{}_{2}^{ij}\Big)\bigg]\nonumber \\
 & \ \ -\frac{1}{6\sqrt{3}}e^{\frac{\varphi}{\sqrt{3}}}\chi\epsilon^{ijk}\bigg[\chi\tilde{F}_{1,jk}\Big(-\chi e^{\frac{2\varphi}{\sqrt{3}}}\pi_{1,i}+(e^{\frac{2\varphi}{\sqrt{3}}}\chi^{2}+3)\pi_{2,i}\Big)\nonumber \\
 & \ \ \ \ \ \ +3\tilde{F}{}_{2,jk}\Big(-e^{\frac{2\varphi}{\sqrt{3}}}\chi\pi_{1,i}+(e^{\frac{2\varphi}{\sqrt{3}}}\chi^{2}+2)\pi_{2,i}\Big)\bigg],
\end{align}
the momentum constraint $C_{i}$ is 
\begin{align}
C_{i} & =-2\text{D}_{j}\pi_{\ i}^{j}+\tilde{F}_{1,ij}\pi_1^{j}+\tilde{F}
_{2,ij}\pi_2^{j}+\pi_{\varphi}\text{D}_{i}\varphi+\pi_{\chi}\text{D}_{i}\chi,
\end{align}
and the Gauss law constraints are
\begin{align}
\mathcal{C}^{(1)}  =-\text{D}_{i}\pi_1^{i}, \qquad \mathcal{C}^{(2)}  =-\text{D}_{i}\pi_2^{i}.
\end{align}

\section{The Kerr--Newman black hole and its black string uplift}\label{app:BS in KN}
The standard Kerr-Newman solution in Boyer-Lindquist coordinates is
\begin{align}
\mathrm{d}s_4^{2} & =-\frac{\Delta}{\rho}\big(\mathrm{d}t-a(1-x^{2})\mathrm{d}\phi\big)^{2}+\frac{\rho}{\Delta}\mathrm{d}r^{2}+\frac{\rho \mathrm{d}x^{2}}{1-x^{2}}+\frac{1-x^{2}}{\rho}\big(a\mathrm{d}t-(r^{2}+a^{2})\mathrm{d}\phi\big)^{2},\nonumber \\
A_1& =-\frac{qr}{\rho}(\mathrm{d}t-a(1-x^{2})\mathrm{d}\phi),\ \ \Delta=r^{2}-2\mu r+a^{2}+q^{2},\ \ \rho=r^{2}+a^{2}x^{2},
\end{align}
 Here $r_0=\mu+\sqrt{\mu^2-a^2-q^2}$ denotes the outer horizon of KN black hole in the shifted radial coordinate, so that $\mu=\frac{a^{2}+q^{2}+r^{2}_{0}}{2r_{0}}$. We assume $r_0^2>a^2+q^2$ for the nonextremal case. Using the action \eqref{eq:einstein-maxwell-action}, the thermodynamic quantities are
\begin{align}
T^{(4)}&=\frac{r^{2}_{0}-a^{2}-q^{2}}{4\pi r_{0}(a^{2}+r^{2}_{0})},\ \ \Omega^{(4)}_{\phi}=\frac{a}{a^{2}+r^{2}_{0}},\ \ \Phi^{(4)}_{{\rm E}}=\frac{qr_{0}}{a^{2}+r^{2}_{0}}, \cr
S^{(4)}&=\frac{\pi(a^{2}+r^{2}_{0})}{G_{4}},\ \ J^{(4)}_{\phi}=\frac{a\mu}{G_{4}},\ \ Q^{(4)}_{{\rm E}}=\frac{q}{G_{4}},\ \ M^{(4)}=\frac{\mu}{G_{4}},
\label{eq:kn-leading-thermodynamics}
\end{align}
and on-shell action satisfies
\begin{equation}
T^{(4)}I_{\rm on-shell}^{(4)}=M^{(4)}-T^{(4)}S^{(4)}-\Omega_{\phi}^{(4)} J_{\phi}^{(4)}-\Phi_{\rm E}^{(4)}Q_{\rm E}^{(4)}
=\frac{\mu-\frac{q^2 r_0}{r_0^2+a^2}}{2G_4}.
\end{equation}

Using the reduction ansatz \eqref{eq:kk-ansatz}, the 4D Kerr-Newman solution is uplifted to the 5D  black string solution
\begin{align}
\mathrm{d}s^{2}_{5} & =\mathrm{d}s^{2}_{4}+(\mathrm{d}z+A_{1})^{2},\ \
 A_{1} =-\frac{qr}{\rho}(\mathrm{d}t-a(1-x^{2})\mathrm{d}\phi),\nonumber \\
 \hat{A}&=A_{2}=\mathrm{d}t(c_{t}-\frac{\sqrt{3}aqx}{r^{2}+a^{2}x^{2}})+\mathrm{d}\phi(-c_{\phi}+\frac{\sqrt{3}q(a^{2}+r^{2})x}{r^{2}+a^{2}x^{2}}),\nonumber \\
c_{t}&=\frac{a c_{\phi}}{a^{2}+r^{2}_{0}},\ \   c_{\phi}=\begin{cases}
\sqrt{3}q, & 0<x<1,\\
-\sqrt{3}q, & -1<x<0,
\end{cases}
\label{eq:kn-uplift}
\end{align}
Here $\hat A=A_2$ denotes the Maxwell one-form, whereas $A_1$ is the KK vector. The solution above is related to the one given in \cite{Compere:2010fm}
by the following replacement of coordinates and integration constants
\begin{equation}
t\rightarrow -t,\quad r\rightarrow r+2ms^{2}_{\beta},\quad q=2mc_{\beta}s_{\beta},\quad\mu=m(1+2s^{2}_{\beta}).
\end{equation}

The constants $c_t,c_\phi$ make the magnetic potential regular in its respective hemispherical patch. Therefore, when calculating the action, we must pay attention to the upper and lower limits of the integral.
From the 5D point of view , the thermodynamic quantities of the uplifted Kerr-Newman black hole are given by
\begin{align}
X^{(5)}&=X^{(4)},\ \ X\in\{T,S,\Omega,J,M\},\cr
\Phi_{{\rm M}}^{(5)}	&=\sqrt{\text{3}}v_{z}^{(5)}=\sqrt{\text{3}}\Phi_{{\rm E}}^{(4)},\ \ Q_{{\rm M}}^{(5)}=\sqrt{3}P_{z}^{(5)}=\frac{\sqrt{3}}{4}Q_{{\rm E}}^{(4)},
\label{eq:kn-uplift-thermodynamics}
\end{align}
where we have used $G_5=G_4 (2\pi R)$. 
Note also the equality 
\begin{equation}\label{eq:kn-work-terms}
v_{z}^{(5)}P_{z}^{(5)}+\Phi_{\rm M}^{(5)} Q_{\rm M}^{(5)}
=\Phi_{\rm E}^{(4)} Q_{\rm E}^{(4)}.
\end{equation}
Therefore, 5D on-shell action differs from the 4D on-shell action as
\begin{equation}
    T I^{(5)}_{\rm on-shell}=T \tilde{I}^{(4)}_{\rm on-shell}+\Phi_{\rm M}^{(5)}Q_{\rm M}^{(5)}.
    \label{eq:45Action}
\end{equation}
The reason is that in truncating the action to the minimal case, we have implicitly performed electromagnetic duality on $A_2$ by the standard procedure with insertion of proper Lagrange  multiplier term in the action, which evaluates to $-\beta\Phi_{\rm M}^{(5)}Q_{\rm M}^{(5)}$.

\section{Reduction of the higher derivative terms in 5D}
\label{app:mixed-chern-simons-reduction}
We give the intermediate steps leading to \eqref{eq:hd-reduction-identity} and \eqref{eq:hd-4d-density}. Cartan's torsion-free structure equation gives
\begin{align}
\hat{\omega}^{ab}&=\omega^{ab}
-\frac12F_{1}^{ab}\hat{e}^{z}, \qquad
\hat{\omega}^{az}=-\frac12F_{1}{}^{a}{}_{b}e^{b},
\label{eq:hd-spin-connection}
\end{align}
Consequently, the independent curvature components are \cite{Banerjee:2011ts}
\begin{align}
\hat{R}^{ab}{}_{cd}
&=R^{ab}{}_{cd}-\frac12F_{1}^{ab}F_{1,cd}
-\frac12F_{1}{}^{[a}{}_{c}F_{1}{}^{b]}{}_{d},\nonumber\\
\hat{R}^{ab}{}_{cz}&=-\frac12\nabla_c F_{1}^{ab}, \ \ 
\hat{R}^{az}{}_{bc}=-\frac12\nabla^a F_{1,bc}, \  \
\hat{R}^{az}{}_{bz}=\frac14F_{1}^{ac}F_{1,bc},   \cr 
\hat{R}^{ab}&=\mathrm{d}\hat{\omega}^{ab}+\hat{\omega}^{a}_{\ c}\land\hat{\omega}^{cb}+\hat{\omega}^{a}_{\ z}\land\hat{\omega}^{zb},\cr \hat{R}^{az}&=-\frac{1}{2}\nabla_{c}(F^{a}_{1,b})e^{c}\land e^{b}-\frac{1}{4}F^{a}_{1,c}F^{c}_{1,b}e^{b}\land\hat{e}^{z},
\label{eq:hd-curvature-reduction}
\end{align}
For the first term, using $\mathrm{d}F_1=0$ and integrating the derivative terms by parts, the Riemann-square term becomes
\begin{align}
\hat{R}_{ABCD}\hat{R}^{ABCD}
={}&R_{abcd}R^{abcd}-\frac12R_{abcd}F_{1}^{ab}F_{1}^{cd}
-2R_{ab}F_{1}^{ac}F_{1}{}^{b}{}_{c}\nonumber\\
&+\frac38(F_{1}^{2})^{2}+\frac58{\rm Tr}F_{1}^{4}
+2\nabla_aF_{1}^{ac}\nabla_bF_{1}{}^{b}{}_{c}
+\nabla_c V^{c},  
\label{eq:hd-riemann-square}
\end{align}
with
\begin{equation}\label{eq:hd-divergence-current}
    V^{c}={}F_{1,ab}\nabla^{c}F_{1}^{ab}-2F_{1}^{ac}\nabla^{b}F_{1,ab}.
\end{equation}
The other parity-even terms reduce to 
\begin{align}
-\frac12\hat{R}_{ABCD}\hat{F}^{AB}\hat{F}^{CD}
={}&\frac32R_{abcd}F_{1}^{ab}F_{1}^{cd}
-6R_{ab}F_{1}^{ac}F_{1}{}^{b}{}_{c}
+\frac32RF_{1}^{2}-\frac{15}{8}(F_{1}^{2})^{2}
+\frac{15}{4}{\rm Tr}F_{1}^{4},\ \nonumber\\
\frac{5}{36}(\hat{F}^{2})^{2}
={}&\frac54(F_{1}^{2})^{2}, \ \ \ \ 
-\frac{13}{24}{\rm Tr}\hat{F}^{4}
={}-\frac{39}{8}{\rm Tr}F_{1}^{4}.
\label{eq:hd-parity-even-reduction}
\end{align}
The total parity even terms are
\begin{align}
  &  	\hat{R}_{ABCD}\hat{R}^{ABCD}-\frac{1}{2}\hat{R}_{ABCD}\hat{F}^{AB}\hat{F}^{CD}+\frac{5}{36}(\hat{F}^{2})^{2}-\frac{13}{24}{\rm Tr}\hat{F}^{4} \cr
=&R_{abcd}R^{abcd}+\frac{3}{2}RF^{2}_{1}-8R_{ab}F^{ac}_{1}F_{1}{}^{b}{}_{c}+R_{abcd}F^{ab}_{1}F^{cd}_{1} \cr
	&-\frac{1}{4}(F^{2}_{1})^{2}-\frac{\text{1}}{2}{\rm Tr}F^{4}_{1}+2\nabla_{a}F^{ac}_{1}\nabla_{b}F_{1}{}^{b}{}_{c}+\nabla_{c}V^{c},
\end{align}

Next, we will reduce the parity odd term and  give the intermediate steps. The parity odd term is
\begin{align}
-\frac{1}{2\sqrt{3}}\hat{\epsilon}^{EABCD}\hat{A}_{E}\hat{R}_{AB}{}^{MN}\hat{R}_{CDMN}\star_5 \boldsymbol 1& =-\frac{2}{\sqrt{3}}\hat{A}\land{\rm Tr}(\hat{R}\wedge\hat{R})\nonumber \\
 & =-\frac{2}{\sqrt{3}}F_{2}\wedge\widehat{{\rm CS}}+\frac{2}{\sqrt{3}}\mathrm{d}(A_{2}\wedge\widehat{{\rm CS}}),
\label{eq:hd-cs-integration}
\end{align}
with
\begin{equation}
\mathrm{d}\widehat{{\rm CS}}={\rm Tr}(\hat{R}\wedge\hat{R}),\  \ \
\widehat{{\rm CS}}=\hat{\omega}^{A}_{\ B}\land\hat{R}^{B}_{\ A}-\frac{1}{3}\hat{\omega}^{A}_{\ B}\land\hat{\omega}^{B}_{\ C}\land\hat{\omega}^{C}_{\ A},
\end{equation}
For a component evaluation one may use
\begin{align}
\hat{\omega}^{A}{}_{B\alpha} & =\hat{\Gamma}^{\gamma}_{\beta\alpha}\hat{e}^{A}_{\ \gamma}\hat{e}^{\ \beta}_{B}-\hat{e}^{\ \beta}_{B}\partial_{\alpha}\hat{e}^{A}_{\ \beta},\nonumber \\
\widehat{{\rm CS}}_{\alpha\beta\gamma} & =3!\big(\frac{1}{2}\hat{\omega}^{M}_{N[\alpha}\hat{R}^{N}_{\ M|\beta\gamma]}-\frac{1}{3}\hat{\omega}^{M}_{\,N[\alpha}\hat{\omega}^{N}_{\,|P|\beta}\hat{\omega}^{P}_{\,|M|\gamma]}\big).
\end{align}
The Chern--Simons three-form can be written as
\begin{align}
\widehat{{\rm CS}}
={}&\hat\omega^{ab}\wedge\hat R_{ba}
+2\hat\omega^{az}\wedge\hat R_{za}
-\frac13\hat\omega^{a}{}_{b}\wedge\hat\omega^{b}{}_{c}
\wedge\hat\omega^{c}{}_{a}
-\hat\omega^z{}_a\wedge\hat\omega^a{}_b\wedge\hat\omega^b{}_z.
\end{align}
Substituting the curvature forms \eqref{eq:hd-curvature-reduction} combines the first and last terms
into the connection on the 4D Lorentz bundle:
\begin{align}
F_2\wedge\widehat{{\rm CS}}
={}&F_2\wedge\hat\omega^{ab}\wedge
\left(\mathrm{d}\hat\omega_{ba}+\hat\omega_{bc}\wedge\hat\omega^c{}_a\right)
+2F_2\wedge\hat\omega^{az}\wedge\hat R_{za}\nonumber\\
&-\frac13F_2\wedge\hat\omega^a{}_b\wedge
\hat\omega^b{}_c\wedge\hat\omega^c{}_a.
\label{eq:app-chern-simons-decomposition}
\end{align}
Only its component supported entirely on the 4D base vanishes, the component containing d$z$ must be retained. The first and third terms give
\begin{align}
&F_2\wedge\hat\omega^{ab}\wedge
\left(\mathrm{d}\hat\omega_{ba}+\hat\omega_{bc}\wedge\hat\omega^c{}_a\right)
\nonumber\\
&\qquad=\frac12F_2\wedge F_{1,ab}R^{ab}\wedge \mathrm{d}z
-\frac12F_2\wedge\omega_{ab}\wedge
\nabla_cF_1^{ba}e^c\wedge \mathrm{d}z
-\frac14F_1^2 F_2\wedge F_1\wedge \mathrm{d}z,\cr
&-\frac13F_2\wedge\hat\omega^a{}_b\wedge
\hat\omega^b{}_c\wedge\hat\omega^c{}_a
=-\frac12F_2\wedge\omega^a{}_c F_1^c{}_b
\wedge\omega^b{}_a\wedge \mathrm{d}z.
\end{align}
Using $\mathrm{d}F_2=0$, the connection-dependent pieces are rearranged by
\begin{align}
\frac12 \mathrm{d}\big(F_2\wedge\omega^{ab}F_{1,ab}\wedge \mathrm{d}z\big)
={}&\frac12F_2\wedge R^{ab}F_{1,ab}\wedge \mathrm{d}z
-\frac12F_2\wedge\omega^a{}_c\wedge\omega^{cb}
F_{1,ab}\wedge \mathrm{d}z\nonumber\\
&-\frac12F_2\wedge\omega^{ab}\wedge \mathrm{d}F_{1,ab}\wedge \mathrm{d}z.
\end{align}
Thus
\begin{align}
F_2\wedge\widehat{{\rm CS}}
+\frac12 \mathrm{d}\big(F_2\wedge\omega^{ab}F_{1,ab}\wedge \mathrm{d}z\big)
=&\big(F_2\wedge F_{1,ab}R^{ab}
-\frac14F_1^2F_2\wedge F_1\big)\wedge \mathrm{d}z
\nonumber\\
&+2F_2\wedge\hat\omega^{az}\wedge\hat R_{za}.
\label{eq:app-chern-simons-reduction}
\end{align}
Using \eqref{eq:F2F1}
$F_2=\sqrt3\star_4 F_1$, we find three contributions
\begin{align}
-\frac{2}{\sqrt3}F_2\wedge F_{1,ab}R^{ab}\wedge \mathrm{d}z
&=-R_{abcd}F_1^{ab}F_1^{cd}\star_4 \boldsymbol{1}\wedge \mathrm{d}z,\nonumber\\
\frac{1}{2\sqrt3}F_1^2F_2\wedge F_1\wedge \mathrm{d}z
&=\frac14(F_1^2)^2\star_4 \boldsymbol{1}\wedge \mathrm{d}z,\nonumber\\
-\frac{4}{\sqrt3}F_2\wedge\hat\omega^{az}\wedge\hat R_{za}
&=\frac12{\rm Tr}F_1^4\star_4 \boldsymbol{1}\wedge \mathrm{d}z.
\end{align}
Finally, we obtain the reduction of parity odd term
\begin{align}
-\frac{2}{\sqrt{3}}\hat{A}\land{\rm Tr}(\hat{R}\wedge\hat{R})
={}&\Big(-R_{abcd}F_{1}^{ab}F_{1}^{cd}
+\frac14(F_{1}^{2})^{2}+\frac12{\rm Tr}F_{1}^{4}\Big)
\star_4 \boldsymbol 1 \wedge \mathrm{d}z\nonumber\\
&+\frac{1}{\sqrt{3}}\mathrm{d}\Big(F_{1,ab}F_{2}\wedge\omega^{ab}_{4}\wedge \mathrm{d}z+2A_{2}\wedge\widehat{{\rm CS}}\Big).
\label{eq:hd-mixed-chern-simons-reduction}
\end{align}
The $R_{abcd}F_{1}^{ab}F_{1}^{cd}$ and $F_1^4$ terms cancel between~\eqref{eq:hd-riemann-square}, \eqref{eq:hd-parity-even-reduction} and~\eqref{eq:hd-mixed-chern-simons-reduction}, we find the following relation
\begin{equation}
\mathcal{L}_{4\partial}^{(5)}\star_{5}\boldsymbol{1}=  \mathcal{L}_{4\partial}^{(4)}\star_{4}\boldsymbol{1}\wedge \mathrm{d}z+\mathcal{L}^{{\rm TD}}_{4\partial},
\end{equation}
where
\begin{align}
\mathcal{L}_{4\partial}^{(4)}= & R_{abcd}R^{abcd}+\frac{3}{2}RF^{2}_{1}-8R_{ab}F^{ac}_{1}F_{1}{}^{b}{}_{c}+2\nabla_{a}F^{ac}_{1}\nabla_{b}F_{1}{}^{b}{}_{c}.\nonumber\\ 
\mathcal{L}^{{\rm TD}}_{4\partial}= & \nabla_{c}V^{c}\star_{4}\boldsymbol{1}\wedge \mathrm{d}z+\frac{1}{\sqrt{3}}\mathrm{d}\big(F_{1,ab}F_{2}\wedge\omega^{ab}_{4}\wedge \mathrm{d}z+2A_{2}\wedge\widehat{{\rm CS}}\big),
\end{align}
These are local identities in the gauge and Lorentz patches.
The three-form $\widehat{{\rm CS}}$ depends on the chosen Lorentz
frame, and $A_2$ is not globally defined for nonzero magnetic charge.
The exact forms must therefore be integrated patch by patch,
including the interface terms, rather than discarded as globally
trivial.


\end{document}